\documentclass[10pt,journal,compsoc]{IEEEtran}
\ifCLASSOPTIONcompsoc
  \usepackage[nocompress]{cite}
\else
  \usepackage{cite}
\fi

\ifCLASSINFOpdf
\else
\fi
\RequirePackage{amsthm}
\usepackage{graphicx}%
\usepackage{multirow}%
\usepackage{amsmath,amssymb,amsfonts}%
\usepackage{amsthm}%
\usepackage{mathrsfs}%
\usepackage{xcolor}%
\usepackage{textcomp}%
\usepackage{manyfoot}%
\usepackage{booktabs}%
\usepackage{algorithm}%
\usepackage{algorithmicx}%
\usepackage{algpseudocode}%
\usepackage{listings}%
\usepackage{bm}
\newtheorem{proposition}{Proposition}
\newtheorem{remark}{Remark}

\begin{document}
%
\title{Modeling Product Ecosystems}
%
%
%
%
\author{Tridib~Banerjee
\IEEEcompsocitemizethanks{\IEEEcompsocthanksitem T. Banerjee was with the Department
of Atmospheric Sciences, Goethe University Frankfurt, Germany, 60348.\protect\\
E-mail: banerjee@iau.uni-frankfurt.de}}

\IEEEtitleabstractindextext{%
\begin{abstract}
This paper develops a dynamical-systems framework for modeling influence propagation in product adoption networks, formulated as a positive linear system with Metzler interaction matrices and utility-based decay. Exact solutions are derived for constant, piecewise-constant, and fully time-varying interaction structures using matrix exponentials and the Peano--Baker series. It establishes five results: (i) positive interactions guarantee nonnegative amplification, (ii) perceived utility saturates after $\approx\!3$ complementary additions (Weber--Fechner), (iii) frequency of comparable introductions dominates incremental quality improvements, (iv) reinforcing interactions yields monotone gains while decay control gives ambiguous effects, and (v) long-run retention under SIS-type dynamics is bounded by the inverse spectral radius of the adoption graph. These results extend epidemic-threshold theory and positive-systems analysis to networked adoption, yielding explicit, calibratable expressions for influence dynamics on networks.
\end{abstract}

}

\maketitle

\IEEEdisplaynontitleabstractindextext

%
\IEEEpeerreviewmaketitle

\section{Introduction}\label{sec:introduction}
Network diffusion and influence propagation arise broadly in epidemics, information spread, and technology adoption. Classical SIS/SIR models characterize thresholds via spectral properties of the contact network, linking persistence to the largest eigenvalue of the adjacency matrix (e.g., \cite{Wang2003,Van_Mieghem_Omic_Kooij_2009}). In the context of \emph{product ecosystems}, rigorous models that combine interaction-driven amplification, cost/utility-based decay, and explicit time variation remain scarce (cf.\ \cite{carrington-2005,sawy-2012,SHANG2022495,shrivastav-2024,robins-2001}). This paper proposes and analyzes a tractable networked dynamical model for ecosystem influence. Section~\ref{sec:theory} presents the model and its properties (positivity, stability, boundedness), and derives exact solutions for constant, piecewise-constant, and fully time-varying interactions via matrix exponentials and the Peano--Baker series \cite{Blanes_Casas_Oteo_Ros_2009}. Section~\ref{sec:Prop} develops implications: amplification guarantees, Weber--Fechner saturation, frequency-vs-quality scaling, control trade-offs (interaction reinforcement vs.\ decay control), and a retention bound via the spectral radius (SIS-like threshold). Section~\ref{sec:Conc} concludes with synthesis, limitations, and directions for empirical calibration.
\section{Theory}\label{sec:theory}
There exists several modeling approaches within different branches of mathematics that maybe adopted to model the influence of a product ecosystem - Graph Neural Networks \cite{ZHOU202057}, Mutual information theoretics \cite{yang2018}, Linear Aggregations \cite{FORNI1999131}, Laplacians \cite{monasson1999diffusion}, Probabilistic Cascades \cite{LI2021125584}, etc, just to name a few. In this paper, the model proposed is a combination of linear aggregation, graph-based (Metzler) interaction matrices, and the utility approach. Let there be $n$ products. We define the \emph{influence vector} $\alpha(t)=(\alpha_1,\dots,\alpha_n)^\top\in\mathbb{R}^n_{\ge0}$, with entries normalized to lie in $[0,1]$ unless noted. Furthermore, let
\begin{itemize}
    \item $\Lambda_{ij}(t)\ge0$ (for $i\ne j$) quantify how product $j$ enhances product $i$, i.e. the interaction term.
    \item $\delta_i(C_i)>0$ quantify the decay/penalty increase with cost/effort $C_i$. let it be defined on the diagonal as a self--leak term.
    \item $\bm{u}(t)\ge0$ quantify the external input, i.e., the exogenous push (marketing/PR/advertising).
\end{itemize}
we then define the \emph{interaction matrix} $M(t)\in\mathbb{R}^{n\times n}$
\begin{equation*}
M_{ij}(t)=
\begin{cases}
\Lambda_{ij}(t), & i\ne j,\\
-\delta_i(C_i), & i=j.
\end{cases}
\end{equation*}
such that the continuous--time model is then,
\begin{equation}\label{eq:ct_model}
\dot\alpha(t)=M(t)\,\alpha(t)+\bm{u}(t),\quad \alpha(t_0)=\alpha_0
\end{equation}
If $\alpha_i$ denotes the net influence of a product $i$, then the above expression suggests that this influence will change over time due to other products within its ecosystem $\alpha_j$ where $j\ne i$. Furthermore, this influence will change positively (i.e., grow) or negatively (i.e., decay) based on the parameters $\Lambda$,$C$, and $u$. Here $\Lambda_{ij}$ represents the typical network weights of linear aggregation between products $i$ and $j$, i.e. how well they enhance each other's experience. $C$ meanwhile is a typical cost penalty function from an utility approach and hence helps to negatively penalize an ecosystem for needing to purchase additional expensive products by the consumer to get the most out of it.
\begin{remark}
(Relation to Graph Laplacian) Our generator $M(t)=\Lambda(t)-\mathrm{diag}(\delta)$ has non-negative off-diagonals (Metzler, see \cite{Shorten_Wirth_Mason_Wulff_King_2007}\cite{Rantzer_2015b}). It is not a graph Laplacian in general. In the special case $\delta_i=\sum_{j\ne i}\Lambda_{ij}$, we have $M=\Lambda-D=-(D-\Lambda)=-L$, i.e., the negative Laplacian where $D$ is the degree matrix. In this paper $\delta_i$ is cost-dependent and need not equal a degree, so $M$ is typically Metzler but not Laplacian.
\end{remark}
\subsection{Properties: Positivity, Stability, Boundedness}
\textit{Positivity.}
If $M(t)$ is Metzler and $u(t)\ge 0$, then each boundary face $\{\alpha_i=0\}$ has $\dot\alpha_i(t)=\sum_{j\ne i} M_{ij}(t)\alpha_j(t) + u_i(t)\ \ge\ 0$, so the non-negative orthant is forward invariant (Nagumo/Kamke). Hence, $\alpha(t)\ge 0$ for all $t\ge t_0$ whenever $\alpha(t_0)\ge 0$.\\
\noindent\textit{Stability.}
For time--invariant $M$, the origin is exponentially stable if $M$ is \emph{Hurwitz} (all eigenvalues have negative real part), i.e., $e^{Mt}\to0$ as $t\to\infty$. With constant input $\bm{u}$, the equilibrium is $-M^{-1}\bm{u}$ (if $M$ is invertible), i.e., $\alpha\to-M^{-1}\bm{u}$ as $t\to\infty$. For time--varying cases, uniform exponential stability can be certified by common Lyapunov functions or boundedness of the transition matrix.

\noindent\textit{Boundedness (nonlinear variant).} To prevent unbounded growth with strong complementarities, a saturating model keeps $\alpha_i\in[0,1]$.
\begin{equation}
\dot\alpha_i=(1-\alpha_i)\sum_{j\ne i}\Lambda_{ij}\alpha_j-\delta_i\alpha_i-\sum_{j\ne i}c_{ij}\alpha_i\alpha_j
\label{eq:saturating}
\end{equation}
Here $c_{ij}$ can be thought of as a crowd penalty term. This preserves non-negativity and caps influence. In many models, only logistical saturation $(1-\alpha_i)$ and self-decay $-\delta_i\alpha_i$ is enough to keep $\alpha_i$ bounded to $[0,1]$ but here, cross-crowding can still unrealistically influence growth. This is why the non-linear decay $c_{ij}$ is further necessary. It is best interpreted as crowding penalties (or competition coefficients) between product i and product j. Economically, it models competition for attention, budget, or screen time.

\subsection{Exact Solutions}
We present the exact solutions in three levels: (i) constant $M$, (ii) piecewise--constant $M$, (iii) fully time--varying $M(t)$.
\subsubsection{Scalar warm-up (intuition)}
For one product, $\dot x=ax+u(t)$, $x(t_0)=x_0$. Multiplying by $e^{-at}$,
\begin{equation*}
\frac{d}{dt}\big(e^{-at}x(t)\big)=e^{-at}u(t)\;
\end{equation*}
or,
\begin{equation*}
 e^{-at}x(t)-e^{-at_0}x_0=\int_{t_0}^{t}e^{-as}u(s)\,ds
\end{equation*}
Hence
\begin{equation}
 \boxed{x(t)=e^{a(t-t_0)}x_0+\int_{t_0}^{t}e^{a(t-s)}u(s)\,ds}
 \label{eq:scalar_solution}
\end{equation}

\subsubsection{Constant $M$ (matrix integrating factor)}
For this case, we then assume $M(t)\equiv M$. Using $\tfrac{d}{dt}e^{-Mt}=-Me^{-Mt}$, multiplying \eqref{eq:ct_model} by $e^{-Mt}$,
\begin{equation*}
\frac{d}{dt}\big(e^{-Mt}\alpha(t)\big)=e^{-Mt}\bm{u}(t)
\end{equation*}
Integrating from $t_0$ to $t$,
\begin{equation*}
 e^{-Mt}\alpha(t)-e^{-Mt_0}\alpha_0=\int_{t_0}^{t}e^{-Ms}\bm{u}(s)\,ds
 \end{equation*}
Multiply by $e^{Mt}$,
\begin{equation}
\alpha(t)=e^{M(t-t_0)}\alpha_0+\int_{t_0}^{t}e^{M(t-s)}\bm{u}(s)\,ds
\label{eq:duhamel}
\end{equation}
This is Duhamel's formula. If $\bm{u}(s)\equiv\bm{u}_0$ and $M$ invertible,
\begin{equation}
\int_{t_0}^{t}e^{M(t-s)}\,ds=\int_{0}^{\Delta t}\!e^{M\tau}d\tau=M^{-1}(e^{M\Delta t}-I),\quad \Delta t=t-t_0
\label{eq:int_exp}
\end{equation}
so
\begin{equation}
\boxed{
\alpha(t)=e^{M\Delta t}\alpha_0+M^{-1}(e^{M\Delta t}-I)\bm{u}_0}
\label{eq:const_input_solution}
\end{equation}

\subsubsection{Piecewise-constant $M$ (interval recursion)}
For this case, we can then partition $[t_0,T]$ into intervals $[t_k,t_{k+1}]$ of length $\Delta t_k$, with $M(t)=M_k$, $\bm{u}(t)=\bm{u}_k$ on each. Applying \eqref{eq:duhamel} on $[t_k,t_{k+1}]$,
\begin{equation}
\alpha(t_{k+1})=e^{M_k\Delta t_k}\alpha(t_k)+\int_{t_k}^{t_{k+1}}e^{M_k(t_{k+1}-s)}\,\bm{u}_k\,ds\label{eq:pc_step}
\end{equation}
With the change of variables $\tau=t_{k+1}-s$, the integral becomes $\int_0^{\Delta t_k}e^{M_k\tau}d\tau\,\bm{u}_k$. If $M_k$ is invertible,
\begin{equation}
\boxed{\alpha(t_{k+1})=e^{M_k\Delta t_k}\alpha(t_k)+M_k^{-1}\big(e^{M_k\Delta t_k}-I\big)\bm{u}_k}
\label{eq:pc_update}
\end{equation}
Chaining (time--ordered) across all intervals yields $\alpha(T)$.

\subsubsection{Fully time-varying $M(t)$ (state-transition matrix)}
In the case of constant $M$, the matrix $e^{M(t-t_0)}$ maps the initial state $\alpha_0$ to $\alpha(t)$. However, if $M$ varies arbitrarily time, this expression can no longer be followed as matrix multiplication in general does not commute. For this, we define the \emph{state--transition matrix} $\Phi(t,t_0)$ which maps a state at time $t_0$ to the state at time $t$. It satisfies the matrix initial value problem,
\begin{equation}
\frac{d}{dt}\Phi(t,t_0)=M(t)\Phi(t,t_0),\quad \Phi(t_0,t_0)=I\label{eq:stm_def}
\end{equation}
and solves the homogeneous system $\dot{\bm{x}}=M(t)\bm{x}$. By variation of constants, we then seek $\alpha(t)=\Phi(t,t_0)\bm{c}(t)$ for some vector $\bm{c}(t)$. Differentiating and matching terms gives $\dot{\bm{c}}(t)=\Phi(t_0,t)\bm{u}(t)$ since $\Phi(t_0,t)=\Phi(t,t_0)^{-1}$, so
\begin{equation*}
\bm{c}(t)=\alpha_0+\int_{t_0}^{t}\Phi(t_0,s)\bm{u}(s)\,ds
\end{equation*}
Using the composition rule $\Phi(t,t_0)\Phi(t_0,s)=\Phi(t,s)$, we obtain the exact solution,
\begin{equation}
\boxed{
\alpha(t)=\Phi(t,t_0)\,\alpha_0+\int_{t_0}^{t}\Phi(t,s)\,\bm{u}(s)\,ds}
\label{eq:tv_solution}
\end{equation}
The transition matrix admits \emph{Peano--Baker series} (time--ordered exponential see \cite{Blanes_Casas_Oteo_Ros_2009}),
\begin{equation}
\Phi(t,t_0)=I+\int_{t_0}^{t}M(\tau_1)d\tau_1+\int_{t_0}^{t}M(\tau_1)\!\int_{t_0}^{\tau_1}\!M(\tau_2)d\tau_2\,d\tau_1+\cdots\label{eq:peano_baker}
\end{equation}
which is necessary since $M(\tau)$ need not always commute. But if $M(\tau_1)$ and $M(\tau_2)$ commute for all times, then
\begin{equation}
\Phi(t,t_0)=\exp\!\Big(\int_{t_0}^{t}M(\tau)\,d\tau\Big),\quad \Phi(t,s)=\exp\!\Big(\int_{s}^{t}M(\tau)\,d\tau\Big)\label{eq:commuting_case}
\end{equation}
giving,
\begin{equation}
\boxed{\alpha(t)=\exp\!\Big(\int_{t_0}^{t}M(\tau)\,d\tau\Big)\alpha_0+\int_{t_0}^{t}\exp\!\Big(\int_{s}^{t}M(\tau)\,d\tau\Big)\bm{u}(s)\,ds}
\label{eq:tv_solution_full}
\end{equation}
It is readily verifiable that for $M(t)=M$ or $M(t)=M_k$, i.e., $M$ equals to constant or piecewise-constant, one readily recovers \eqref{eq:const_input_solution} or time-chained \eqref{eq:pc_update} respectively.
\subsection{Discrete-Time Form and Estimation}
Sampling at period $\Delta t$ and assuming $M$ and $\bm{u}$ are held over each step, \eqref{eq:pc_update} yields
\begin{equation}
\alpha_{t+1}=A_t\,\alpha_t+B_t\,\bm{u}_t,\quad
A_t=e^{M_t\Delta t},\quad B_t=\int_0^{\Delta t}e^{M_t(\Delta t-\tau)}\,d\tau\label{eq:dt_model}
\end{equation}
If $M_t$ invertible, $B_t=M_t^{-1}(e^{M_t\Delta t}-I)$. This linear regression form supports constrained estimation (e.g., $A_t$ Metzler--consistent), and recovering $M_t\approx \tfrac{1}{\Delta t}\log A_t$ when appropriate. Regularization (e.g., sparsity on off--diagonals) improves identifiability.

\section{Propositions}\label{sec:Prop}
This section presents several numerical studies to further diagnose the viability of the proposed theory and draw valuable insights from it.
We analyze canonical scenarios using the exact solutions. We study the system,
\begin{equation}
  \dot\alpha(t)=M(t)\alpha(t)+\bm{u}(t),\quad 
  M(t)=\Lambda(t)-\mathrm{diag}(\delta_1,\ldots,\delta_n),
\end{equation}
with $\alpha(t)\in\mathbb{R}^n_{\ge0}$, $\Lambda_{ij}(t)\ge 0$ ($i\neq j$), $\delta_i>0$, $\bm{u}(t)\ge 0$.  
To isolate the ecosystem effect, let $\alpha^\circ(t)$ denote the trajectory of the \emph{decoupled} system with $\Lambda\equiv 0$ and same $\delta,u$. We then define the \emph{amplification factor},
\begin{equation}
\mathcal{A}_i(t)=\frac{\alpha_i(t)}{\alpha_i^\circ(t)}
\end{equation}

\subsection{Amplification under Constant Interactions}\label{prop1}
\begin{proposition}
If $\Lambda$ is constant with $\Lambda_{ij}\ge 0$ for $i\neq j$, then for all $t\ge t_0$,
$\alpha_i(t)\;\ge\;\alpha_i^\circ(t)\quad\text{and}\quad\mathcal{A}_i(t)\ge 1$
whenever $\alpha_i^\circ(t)>0$.  
\end{proposition}
\begin{proof}
For the system dynamics defined above, let $z(t)=\alpha(t)-\alpha^{\circ}(t)$ such that,
\begin{equation}
    \dot{z}(t)=\dot{\alpha}(t)-\dot{\alpha^{\circ}}(t)=M(t)z(t)+\big{(}M(t)-M^{\circ}(t)\big{)}\alpha^{\circ}(t)
\end{equation}
since $\alpha=\alpha^{\circ}+z$ but $M(t)-M^{\circ}(t)$ is just $\Lambda$. Therefore,
\begin{equation}
    \dot{z}(t)=M(t)z(t)+\Lambda\alpha^{\circ}(t),\quad z(t_0)=0
\end{equation}
which is a linear, time-varying system driven by a non-negative $\Lambda\alpha^{\circ}(t)$. Since $M(t)$ is Metzler for all $t$ (non-negative off-diagonals) and $\bm{u}\ge 0$, the flow is positive, hence the state-transition matrix $\Phi(t,s)$ of $M(\cdot)$ satisfies $\Phi(t,s)\ge 0$ entrywise for $t\ge s$, and the baseline trajectory obeys $\alpha^\circ(s)\ge 0$. Hence from (\ref{eq:tv_solution}),
\begin{equation}
z(t)=\int_{t_0}^{t}\Phi(t,s)\,\Lambda\,\alpha^\circ(s)\,ds\ \ge\ 0   
\end{equation}
or $\alpha(t)\ge\alpha^\circ(t)$ componentwise for all $t\ge t_0$. Whenever $\alpha_i^\circ(t)>0$ this yields $\mathcal{A}_i(t)\ge 1$. If in addition some $\Lambda_{ij}>0$ and $\alpha_j^\circ$ is nontrivial, then the integrand has a strictly positive $i$-th component
\begin{equation}
z_i(t) = \int_{t_0}^t \sum_{k=1}^n\sum_{j=1}^n \Phi_{ik}(t,s)\Lambda_{kj}\,\alpha_j^\circ(s)\,ds
\end{equation}
hence a strict amplification $\mathcal{A}_i(t)>1$.
\begin{remark}
The argument does \emph{not} require $\delta_i(t)$ to be constant—only that $M(t)$ remain Metzler and the baseline/input be non-negative. If $\Lambda\equiv 0$ then $\bm{z}\equiv 0$ and $\mathcal{A}_i(t)\equiv 1$; otherwise strict inequality holds under mild excitation of the baseline.
\end{remark}
\end{proof}
\subsection{Steady Identical Growth and Weber--Fechner Saturation}\label{prop2}
\begin{proposition}
With $N$ identical add-ons, each of incremental strength $\beta>0$, perceived utility under the Weber--Fechner law, $p(N)=\kappa \log(1+N\beta)$ has diminishing returns. The saturation point (where the marginal gain halves relative to the first device) occurs at $N^*=\frac{1}{\beta}-1$.
\end{proposition}
\begin{proof}
To study the aggregate effect of adding $N$ comparable complementary products, instead of carrying full matrices, we approximate the net ecosystem-induced boost beyond baseline as linear in $N$ at first order,
\begin{equation}
\Delta \mathcal{A}(N) \approx N \beta
\end{equation}
with $\beta>0$ capturing the per-add-on marginal gain (derived from the off-diagonal $\Lambda_{ij}$ and decay structure, averaged over the horizon). So the amplification factor is approximated by,
\begin{equation}
\mathcal{A}(N) \approx 1+ N \beta
\end{equation}
Now, consumers don't feel $\mathcal{A}$ linearly. As per Weber–Fechner law (see \cite{Stevens_1957}\cite{Teghtsoonian_1971}), perceived magnitude grows logarithmically in the raw stimulus. So if we take raw stimulus as $\mathcal{A}$ or equivalently $1+N\beta$, then the marginal perceived gain is,
\begin{equation}
\frac{dp}{dN}=\frac{\kappa\beta}{1+N\beta}
\end{equation}
At $N=0$, the marginal gain is $\kappa\beta$. Defining $N^*$ by $\tfrac{dp}{dN}|_{N^*}=\tfrac{1}{2}\kappa\beta$ and solving
\begin{equation}
\frac{\kappa\beta}{1+N^*\beta}=\frac{1}{2}\kappa\beta    
\end{equation}
gives $N^*=1/\beta-1$.
\end{proof}
\begin{remark}
Typical $\beta\in(0.25,0.5)$ yields $N^*\in[1,3]$, giving an empirical ``three-device'' rule-of-thumb.
\end{remark}
\subsection{Unsteady Growth -- Frequency vs. Quality}\label{prop3}
\begin{proposition}
If $N_g$ identical add-ons (strength $\beta$) are introduced at each step over $S$ steps, the amplification factor is $\mathcal{A} = 1+\beta N_g \frac{S(S+1)}{2}$. Hence frequency (larger $S$) compounds quadratically, while quality ($\beta$) only scales linearly.
\end{proposition}
\begin{proof}
Consider discrete-time updates (piecewise-constant $M$),
\begin{equation}
\alpha_{s+1}=A\alpha_s+B(u+\beta N_g)
\end{equation}
with $A\approx e^M$, $B=\int_0^1 e^{M(1-\tau)}d\tau$. Unrolling for $S$ steps,
\begin{equation}
\alpha_S=A^S\alpha_0+\sum_{k=0}^{S-1} A^k B u + \Big(\sum_{k=0}^{S-1} A^k B\Big)\beta N_g  
\end{equation}
Then defining the baseline (no add-ons),
\begin{equation}
\alpha_S^o=A^S\alpha_0+\sum_{k=0}^{S-1} A^k B u
\end{equation}
Leading to the amplification factor,
\begin{equation}
\mathcal{A}=1+\frac{\Big(\sum_{k=0}^{S-1} A^k B\Big)\beta N_g}{A^S\alpha_0+\sum_{k=0}^{S-1} A^k B u}
\end{equation}
For stable $A\approx I$ (i.e. small steps/weak per-step dynamics), $\sum_{k=0}^{s-1}A^k\approx s I$. Each add-on then contributes over all remaining steps, giving cumulative effect,
\begin{equation}
\mathcal{A}\approx1+\beta N_g \sum_{s=1}^{S} s=1+\beta N_g \frac{S(S+1)}{2} 
\end{equation}
\end{proof}

\subsection{Optimal Investment -- Interactions vs. Cost Cuts}\label{prop4}
\begin{proposition} Interactions guarantee amplification while cost cuts with weaker integration are ambiguous, i.e., if the cumulative weighted influence is $J(\Lambda,\delta)=\int_{t_0}^{T} w^\top \alpha(t)\,dt$, and the cumulative amplification is $\mathcal{A}_{\mathrm{cum}}
={\int_{t_0}^{T} w^\top \alpha(t)\,dt}/{\int_{t_0}^{T} w^\top \alpha^{\circ}(t)\,dt}$ then, (i) For any perturbation with $\Delta\Lambda_{ij}\ge 0$ and $\Delta\delta\equiv 0$, $\partial \mathcal{A}_{\mathrm{cum}}/\partial \Lambda_{ij}\ge 0$, and (ii) while for one-parameter policy $\eta\mapsto(\Lambda(\eta),\delta(\eta))$ such that $\partial \delta_i/\partial\eta<0$ and $\partial \Lambda_{ij}/\partial\eta\le 0$, then, in general, $d_{\eta}(J,\mathcal{A}_{\mathrm{cum}})$ is ambiguous.
\end{proposition}

\begin{proof}
Given \eqref{eq:ct_model}, we fix non-negative weights $w\ge 0$ and define the cumulative weighted influence as,
\begin{equation}
J(\Lambda,\delta)=\int_{t_0}^{T} w^\top \alpha(t)\,dt
\end{equation}
and the cumulative amplification as,
\begin{equation}
\mathcal{A}_{\mathrm{cum}}
=\frac{\int_{t_0}^{T} w^\top \alpha(t)\,dt}{\int_{t_0}^{T} w^\top \alpha^{\circ}(t)\,dt}
\end{equation}
where $\alpha^{\circ}$ is the baseline trajectory with the \emph{same} $\delta,u$ but $\Lambda\equiv 0$. We then consider a small, constant perturbation $(\Delta\Lambda,\Delta\delta)$ such that $\Delta M=\Delta\Lambda-\mathrm{diag}(\Delta\delta)$.
Let $\tilde{\alpha}$ be the perturbed trajectory and $\Delta\alpha(t)=\tilde{\alpha}(t)-\alpha(t)$.
Subtracting dynamics and keeping first-order terms gives the linear variation,
\begin{equation}
\dot{\Delta\alpha}(t)=M(t)\,\Delta\alpha(t)+\Delta M\,\alpha(t),\quad
\Delta\alpha(t_0)=0
\end{equation}
Let $\Phi(t,s)$ be the state-transition matrix of $M(\cdot)$. Similar to (\ref{eq:tv_solution}), using variation of constants,
\begin{equation}\label{eq:sens}
\Delta\alpha(t)\;=\;\int_{t_0}^{t}\Phi(t,s)\,\Delta M\,\alpha(s)\,ds
\end{equation}
\medskip
\noindent
Because $M(t)$ is Metzler and $\bm{u}\ge 0$, the flow is positive:
\begin{equation}
\Phi(t,s)\ \ge\ 0\ \text{entrywise for }t\ge s,\quad
\alpha(t)\ \ge\ 0\ \text{for all }t\in[t_0,T]
\end{equation}
\medskip
\noindent
By definition,
\begin{equation}
\Delta J\;=\;\int_{t_0}^{T} w^\top \Delta\alpha(t)\,dt
\stackrel{\eqref{eq:sens}}=\int_{t_0}^{T}w^\top\left(\int_{t_0}^{t}
 \Phi(t,s)\,\Delta M\,\alpha(s)\,ds\,\right)dt
\end{equation}
Here, the domain of integration is the triangle $R = \{(s,t): t_0 \le s \le t \le T\}$. Since the integrand is positive ($M(t)$ is Metzler), using Fubini to swap double integrals,
\begin{equation}
\Delta J=\int_{t_0}^{T}\left(\int_{s}^{T}
w^\top \Phi(t,s)\,\,dt\right)\Delta M\,\alpha(s)\,ds
\end{equation}
Defining a non-negative ``downstream value'' vector
\begin{equation}
q(s)\;=\;\int_{s}^{T}\Phi(t,s)^\top w\,dt\ \in \mathbb{R}^n_{\ge 0}
\end{equation}
we then obtain 
\begin{equation}
\Delta J = \int_{t_0}^T q(s)^\top \Delta M \alpha(s)\,ds
\end{equation}
Expanding $q(s)^\top \Delta M \alpha(s) = \sum_{i=1}^n \sum_{j=1}^n \Delta M_{ij}\, q_i(s)\, \alpha_j(s)$ and substituting $M$ back finally gives the clean decomposition
\begin{equation}\label{eq:master}
\Delta J
=\sum_{i\neq j}\Big(\Delta\Lambda_{ij}\int_{t_0}^{T} q_i(s)\,\alpha_j(s)\,ds\Big)
\;-\;\sum_{i}\Big(\Delta\delta_{i}\int_{t_0}^{T} q_i(s)\,\alpha_i(s)\,ds\Big)
\end{equation}
where all integrands are non-negative by step-positivity. For \emph{(i) Pure interaction investment.}
If $\Delta\Lambda_{ij}\ge 0$ and $\Delta\delta\equiv 0$, each term in the first sum of \eqref{eq:master} is $\ge 0$, hence $\Delta J\ge 0$.
Since the baseline in $\mathcal{A}_{\mathrm{cum}}$ uses $\Lambda\equiv 0$, its denominator does not depend on $\Lambda$. Therefore,
\begin{equation}
\frac{\partial \mathcal{A}_{\mathrm{cum}}}{\partial \Lambda_{ij}}
=\frac{1}{\int_{t_0}^{T} w^\top \alpha^{\circ}\,dt}\,
\frac{\partial J}{\partial \Lambda_{ij}}
\ \ge\ 0    
\end{equation}
For \emph{(ii) Cost cut with integration trade-off.}
Let a scalar policy $\eta\mapsto(\Lambda(\eta),\delta(\eta))$ satisfy
$\partial\delta_i/\partial\eta<0$ (lower decay/penalty) and
$\partial\Lambda_{ij}/\partial\eta\le 0$ (weaker integration).
Differentiating \eqref{eq:master} along $\eta$ gives,
\begin{equation}
\frac{dJ}{d\eta}
=\sum_{i\ne j}\!\Big(\frac{\partial\Lambda_{ij}}{\partial\eta}\Big)\!\int q_i\,\alpha_j\,ds
\;-\;\sum_{i}\!\Big(\frac{\partial\delta_i}{\partial\eta}\Big)\!\int q_i\,\alpha_i\,ds   
\end{equation}
The first sum is $\le 0$ while the second (with the minus sign and $\partial\delta_i/\partial\eta<0$) is $\ge 0$.
Hence the sign of $dJ/d\eta$ (and of $d\mathcal{A}_{\mathrm{cum}}/d\eta$) is \emph{not} determined in general, cost cuts are not guaranteed to increase cumulative influence or amplification. A sufficient condition for $dJ/d\eta>0$ is,
\begin{equation}
 \sum_{i}\Big|\tfrac{\partial\delta_i}{\partial\eta}\Big| \int q_i\,\alpha_i\,ds
\;>\;\sum_{i\ne j}\Big|\tfrac{\partial\Lambda_{ij}}{\partial\eta}\Big| \int q_i\,\alpha_j\,ds   
\end{equation}
and the reverse inequality makes $dJ/d\eta<0$. Concluding, part (i) shows interaction investments strictly raise $J$ and $\mathcal{A}_{\mathrm{cum}}$.
Part (ii) shows that cost cuts (via $\delta\downarrow$) coupled with weaker integration ($\Lambda\downarrow$) have ambiguous net effect; they may raise or lower amplification depending on the relative magnitudes of the two terms in \eqref{eq:master}.
\end{proof}

\begin{remark}[Practical ``ROI'' reading]
The factor $\int q_i(s)\alpha_j(s)\,ds$ in \eqref{eq:master} is the marginal value of edge $(j\!\to\! i)$: it multiplies the upstream signal $\alpha_j$ by the downstream sensitivity $q_i$.
With per-edge costs $k_{ij}>0$, the local return-on-investment is
$\mathrm{ROI}_{ij}=\big(\int q_i\alpha_j\,ds\big)/k_{ij}$; interaction budget should prioritize edges with the largest $\mathrm{ROI}_{ij}$, while recognizing that simultaneous cost cuts that weaken $\Lambda$ can offset these gains.
\end{remark}
\subsection{Retention Bounded by Spectral Radius}\label{prop5}
\begin{proposition}
For SIS (Susceptible–Infected–Susceptible) type adoption dynamics, persistence occurs only if adoption pressure $\tau$ is above a certain critical value $\tau_c$ 
\end{proposition}
\begin{remark}
The SIS-type adoption model studied here (see \cite{Wang2003}\cite{Van_Mieghem_Omic_Kooij_2009}) can be viewed as a social-layer analog of our boundedness equation (\ref{eq:saturating}). In (\ref{eq:saturating}), the factor $(1-\alpha_i)$ ensures that adoption of product $i$ cannot exceed 1 (saturation), while the $-\delta_i\alpha_i$ term allows influence to decay. Similarly, in SIS dynamics, each user segment $x_i(t)$ can adopt $(\beta A x)$ or churn $(-\delta x_i)$, with no permanent immunity, i.e., adoption is reversible. Thus, this proposition extends the saturation idea from product–product layer (bounded influence $\alpha_i$) to user–user layer (bounded adoption $x_i$).
\end{remark}
\begin{proof}
Let $x(t)\in[0,1]^N$ denote the adoption fraction in each segment at time $t$. We then consider retainment across a social adoption graph with adjacency $A\in\mathbb{R}^{N\times N}_{\ge 0}$ on $N$ user segments (or individuals aggregated to segments). Users can churn (become susceptible again) at rate $\delta>0$ and can (re)adopt through social exposure at per-edge rate $\beta>0$. A SIS mean-field consistent with our saturation modeling can then be defined as, 
\begin{equation}
\dot{x}(t)\;=\;-\delta\,x(t)\;+\;\beta\,A\,x(t)\;-\;\text{(higher-order saturation terms)}
\end{equation}
The last term gathers the nonlinear crowding/saturation effects (e.g., factors like $x_i(1-x_i)$ or cross-crowding) that keep $x\in[0,1]^N$. For near zero adoption $(x\approx 0)$, those higher-order terms vanish, and the linear part dominates. The effective adoption pressure can then be, \begin{equation}
\tau=\beta/\delta   
\end{equation}
which is dimensionless. For the linear case $x\approx 0$ making the saturating terms $o(\|x\|)$. The dominant dynamics are then,
\begin{equation}
 \dot{x}(t)\;\approx\; \big(\beta A - \delta I\big)\,x(t)   
\end{equation}
The eigenvalues of $\beta A - \delta I$ are $\beta\lambda_i(A)-\delta$ with the largest one being $\beta\lambda_{\max}(A)-\delta$. If 
\begin{equation}
\beta\lambda_{\max}(A)-\delta\le 0\quad \textrm{or,} \quad\beta/\delta\le \frac{1}{\lambda_{\max}(A)}
\end{equation}
then all eigenvalues are $\le 0$ and the zero state is (locally) attractive: any small adoption decays to zero, i.e., no persistence. But if 
\begin{equation}
\beta\lambda_{\max}(A)-\delta> 0\quad \textrm{or,} \quad\beta/\delta> \frac{1}{\lambda_{\max}(A)}
\end{equation}
then the zero state is (locally) unstable, so with the saturating terms bounding $x$ in $[0,1]^N$, a nonzero equilibrium can emerge and be sustained, i.e., persistence possible. Thus the critical adoption pressure is $\tau_c\;=\;1/\lambda_{\max}(A)$ making persitence only possible when,
\begin{equation}
 \tau\;>\;\frac{1}{\lambda_{\max}(A)}\;=\;\tau_c  
\end{equation}
\end{proof}
\begin{remark}
This is the classic epidemic threshold: even with large global base, retainment cannot exceed the bound set by $\lambda_{\max}(A)$.
\end{remark}
\begin{remark}
Earlier, we showed how ecosystem interactions $\Lambda$ amplify product influence $\alpha(t)$. That's product–product synergy. Here, we're adding a social retainment layer across user segments, with graph $A$. This user–user layer determines whether adoption survives churn. Even if $\Lambda$ makes products reinforce each other (larger $\alpha$), long-run retainment cannot exceed the limit set by the social graph: if $\tau\le 1/\lambda_{\max}(A)$, social re-adoption pressure is too weak vs. churn, and adoption eventually dies out.
\end{remark}

\section{Conclusion}\label{sec:Conc}
In this paper, an explicit linear-time model for ecosystem influence is proposed. Exact solutions are derived and influence growth with respect to various ecosystem factors are assessed. Further propositions are thus laid forth and which are re-stated, this time in terms of economic actions, below.
\begin{itemize}
    \item Proposition \ref{prop1}: \textit{For an ecosystem to exert any amplified influence on its products, at least one such product needs to be non-identical and (or) non-independent to the others.}
    \item Proposition \ref{prop2}: \textit{For any given ecosystem, the number of products a consumer needs to own beyond which there is a sharp fall off in perceived additional influence is approximately 3.}
    \item Proposition \ref{prop3}: \textit{For any given ecosystem, if all tertiary products are of comparable value, then the ecosystem influence depends much more on the frequency of purchase than it does on the quality of those products.}
    \item Proposition \ref{prop4}: \textit{For any given ecosystem, the safest investment one can make for network growth is in increasing the interactions between all its products.}
    \item Proposition \ref{prop5}: \textit{For any given nascent ecosystem seeking to gain market-share, the best approach would be to prioritize regional market penetration, even if it comes at the cost of delaying entry into the global market.}
\end{itemize}
Together, these propositions suggest that while ecosystem interactions reliably amplify influence, the perception of value saturates after a small number of products, frequency of tertiary additions dominates over quality, and retention is fundamentally constrained by the social network structure. These insights unify product design, marketing frequency, pricing strategy, and social adoption into a single mathematical framework. Future work could incorporate stochastic shocks, heterogeneous user types, and empirical calibration on real-world ecosystems to validate and refine the model.\\

Finally, the nonlinear crowding coefficients $c_{ij}$ introduced in (\ref{eq:saturating}) are worth recalling. They capture the finite attention or budget that products compete for, and mathematically ensure that adoption levels remain bounded in $[0,1]$. This guarantees that the amplification mechanisms described in Propositions $2-4$ operate within realistic limits rather than diverging indefinitely.\\

In closing remarks, it is also worth restating the limitations of this theory, namely the qualitative nature of parameters like $\alpha^0,\Lambda$. This prevents the proposed theory from being fully deterministic and leaves the readers with the burden of attributing appropriate values themselves. However, with sufficient data on the market and userbase, it should be relatively straightforward to quantify the said parameters.
\section*{Declarations}
The author declares no competing interest.



\bibliographystyle{IEEEtran}
\bibliography{sn-bibliography.bib}
\end{document}